\documentclass[%
 reprint,
superscriptaddress,
 amsmath,amssymb,
 aps,
]{revtex4-2}

\usepackage{graphicx}
\usepackage{svg}
\usepackage{subcaption}
\usepackage{caption}
\usepackage{bicaption}
\usepackage{dcolumn}
\usepackage{bm}
\usepackage{braket} 
\usepackage{float}
\usepackage{hyperref}
\hypersetup{  colorlinks=true, linkcolor=blue, citecolor=red, urlcolor=blue  }
\usepackage{physics}

\usepackage{color,xcolor}

\DeclareMathOperator{\diag}{diag}
\DeclareMathOperator{\tin}{in}
\DeclareMathOperator{\tout}{out}

\begin{document}
\preprint{APS/123-QED}

\title{Quantum Computing by Quantum Walk on Quantum Slide}

\author{Fan Wang}
\thanks{These authors contributed equally to this work}%
\affiliation{%
 Department of Physics, Southern University of Science and Technology, Shenzhen 518055, China
}%
\author{Bin Cheng}%
\thanks{These authors contributed equally to this work}
\affiliation{%
 Department of Physics, Southern University of Science and Technology, Shenzhen 518055, China
}%
\affiliation{%
Shenzhen Institute for Quantum Science and Engineering,
Southern University of Science and Technology, Shenzhen 518055, China.
}%
\affiliation{Centre for Quantum Software and Information, Faculty of Engineering and Information Technology,
University of Technology Sydney, NSW 2007, Australia.}
\author{Zi-Wei Cui}
\email{11849207@mail.sustech.edu.cn}
\author{Man-Hong Yung}
\email{yung@sustech.edu.cn}
\affiliation{%
 Department of Physics, Southern University of Science and Technology, Shenzhen 518055, China
}%
\affiliation{%
Shenzhen Institute for Quantum Science and Engineering,
Southern University of Science and Technology, Shenzhen 518055, China.
}%

\begin{abstract}

Continuous-time quantum walk is one of the alternative approaches to quantum computation, where a universal set of quantum gates can be achieved by scattering a quantum walker on some specially-designed structures embedded in a sparse graph [Childs, Phys. Rev. Lett. 2009]. Recent advances in femtosecond laser-inscribed optical waveguides represent a promising physical platform for realizing this quantum-walk model of quantum computation. However, the major challenge is the problem of preparing a plane-wave initial state. Previously, the idea of quantum slide has been proposed and experimentally realized for demonstrating the working principle of NAND tree [Wang et al. Phy. Rev. Lett. 2020]. Here we show how quantum slide can be further applied to realize universal quantum computation, bypassing the plane-wave requirement. Specifically, we apply an external field to the perfect-state-transfer chain, which can generate a moving Gaussian wave packet with an arbitrary momentum. When the phase is properly tuned, the universal gate set in Childs' proposal can be realized in our scheme. Furthermore, we show that the gate fidelities increase with the length of the slide, and can reach unity asymptotically.

\end{abstract}

\maketitle

\section{Introduction}

Random walk is a powerful classical algorithm for a large class of search and sampling problems. As its quantum counterpart, quantum walk (QW) is an interesting framework for designing fast quantum algorithms in continuous time or discrete time~\cite{magniez2007quantum1,jeffery2013nested,ambainis2004coins, childs2007discrete,childs2012quantum,le2014improved,bernstein2013quantum,ambainis2007quantum,magniez2007quantum2,buhrman2004quantum,le2012improved,dorn2007quantum}. Below, we shall focus on continuous-time QW, which deals with the time evolution of a quantum state, where the Hamiltonian is in the form of an adjacency matrix of a certain graph~\cite{farhi1998quantum}.

Continuous-time quantum walk (CTQW) exploits exotic quantum phenomena to achieve quantum speedup against its classical counterparts.
For instance, utilizing quantum superposition or interference property, it achieves an exponential algorithmic speedup for the black-box traversal problem on the glued tree graph \cite{childs2002example,childs2003exponential}, compared with any classical algorithm (even those not based on a classical random walk). 
Also, quadratic speedup against classical computers (as powerful as Grover's algorithm \cite{grover1996fast}) can be obtained for the spatial search problems on several different graphs \cite{childs2004spatial}.
Furthermore, due to the walker's scattering properties in distinct graph structures, it can evaluate the NAND tree problem faster than ever known best classical algorithms \cite{farhi2007quantum} and is proved to be a universal model for quantum computation \cite{childs2009universal,childs2013universal}.

In the above applications of CTQW, especially those considering scattering processes, preparing initial wave packets with appropriate momentum is an important step. 
For example, in the NAND tree problem, an initial wave packet is required whose momentum is around $-\frac{\pi}{2}$ to start the evaluation process. 
Moreover, in Childs' method, initial wave packets with momentum only around $-\frac{\pi}{4}$ is an essential element to implement universal quantum gates in the CTQW framework~\cite{childs2009universal}.
However, Childs' method uses an infinitely long structure called momentum filter to filter out undesired momentum and prepare an initial wave packet with the desired momentum, which may be infeasible experimentally.
As far as we know, few studies in CTQW have discussed the issue of preparing an initial wave packet that possesses a proper momentum. 
On the other hand, a recent work \cite{wang2020integrated} proposed a method to realize Gaussian wave packets of $-\frac{\pi}{2}$ momentum by adopting a ``quantum slide'' used for solving the NAND tree problem and is experimentally verified.
Unfortunately, the momentum of the wave packets formed in the slide is fixed to $-\frac{\pi}{2}$, which limits its application to other problems.

In this work, based on the quantum slide method, we present a new scheme capable of generating wave packets with arbitrary momentum. 
By adding an external field, a novel Hamiltonian is obtained that can ``accelerate or decelerate'' (which means changing the momentum of) the original wave packets with fixed $-\frac{\pi}{2}$ momentum. 
Furthermore, by adjusting the strength of the external field and the evolution time, the momentum can be regulated precisely and cover the entire momentum space of $(-\pi,\pi]$. 
After showing its capability of carrying out arbitrary momentum, we apply our scheme to simulate universal quantum computation using Childs' method, where preparing wave packets with momentum $-\pi/4$ is essential. 
Our numerical experiments show that we can use relatively short quantum slide to achieve high-precision gates. 
Moreover, we found the precision can be improved by simply increasing the length of the slide and could reach 1 asymptotically when the slide is long enough.

\section{quantum walk and quantum slide}

Continuous-time quantum walk was first introduced in~\cite{farhi1998quantum} by extending the classical Morkov process to the quantum regime.
It can be described as Hamiltonian evolution of the adjacency matrices indicating the connectivity of the underlying graphs or networks.
Given a graph $G$ of $N+1$ nodes and with adjacency matrix $A$, we represent the walker's state in the basis $\{\ket{0},\ket{1},\dots, \ket{N}\}$ and use the time-evolution operator $e^{-iAt}$ to characterize the evolution.
Consider quantum walk in a one-dimensional chain.
In this case, the Hamiltonian or adjacency matrix corresponding to the chain can be described by an $(N+1)$-by-$(N+1)$ Jacobi matrix in the following form:
\begin{equation}\label{eq:general_H}
H =
\begin{pmatrix}
B_0 & J_1 & & & & \\
J_1 & B_1 & J_2 & &  &\\
 & \ddots & \ddots & \ddots &\\
 & & \ddots & \ddots & J_N &\\
 & & & J_N& B_N& \\
\end{pmatrix} \ .
\end{equation}
In physics, such Hamiltonians could be used for depicting the one-excitation subspace of a quantum spin chain with the nearest-neighbor Heisenberg interaction, where $J_n$ is the coupling strength between the $(n-1)$-th site and the $n$-th site and $B_n$ is the strength of the magnetic field at the $n$-th site ($n = 0,1,2\dots, N$)~\cite{vinet2012construct}.

This form of Hamiltonian can be used to describe perfect state transfer (PST) in a pre-engineered coupling chain with $N+1$ nodes~\cite{christandl2004perfect}. 
The Hamiltonian $H_{\rm PST}$ for perfect state transfer can be obtained by setting 
\begin{align}
   J_n &= \sqrt{n(N+1-n)} & B_n &= 0 \ ,
\end{align}
in Eq.~\eqref{eq:general_H}.
Indeed, for the site $\ket{r}$, the transition amplitude of the time evolution governed by $H_{\rm PST}$ at time $t$ ($0 \leq t \leq \frac{\pi}{2}$) is given by~\cite{wang2020integrated},
\begin{equation}\label{eq:H_PST_amplitude}
  \mel{r}{ e^{-i t H_{\rm PST}}}{0}= \sqrt{\binom{N}{r}} (\sin{t})^r (\cos{t})^{N-r} e^{-i\frac{\pi}{2}r} \ .
\end{equation}
Then, by setting $r = N$ and $t = \pi/2$, one can show that 
\begin{equation}
 \mel{N}{ e^{-i\frac{\pi}{2} H_{\rm PST}}}{0} = 1 \ .
\end{equation}
That is, quantum information can be perfectly transferred in the spin chain in an arbitrarily long distance $N$ after a constant time $t =\pi/2$~\cite{christandl2004perfect}.

Besides, with $t = \frac{\pi}{4}$ and $N$ large enough in $H_{\rm PST}$, a Gaussian wave packet with momentum distributed around $-\frac{\pi}{2}$ can be obtained on the pre-engineered chain via the binomial-Gaussian approximation, achieving the quantum slide scheme~\cite{wang2020integrated}.
The basic construction of the quantum slide scheme is to use half of such a pre-engineered chain called ``quantum slide'', and connect a uniformly coupled chain called ``runway'' after it. 
The former is used to generate a wave packet with a specific shape and momentum (cutting in half is mainly to prevent further unwanted evolution), while the latter is used to stably transmit the wave packet in preparation for the subsequent scattering process.

\begin{figure}[t]
    \centering
    \includegraphics[width=\linewidth]{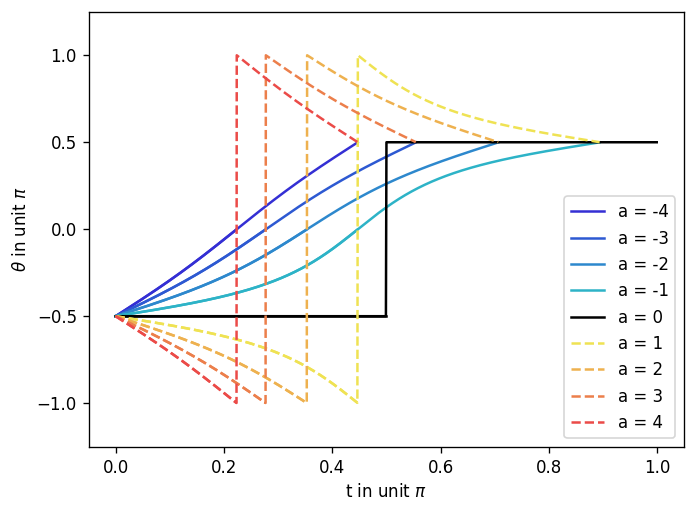}
    \caption{Momentum $\theta$ as a function of evolution time t when a has different values. For each value of a, the momentum in one period of t is depicted.}
    \label{fig:momentum}
\end{figure}

\section{wave packets with arbitrary momentum}

Here, we generalize the Hamiltonian $H_{\rm PST}$ to obtain a new Hamiltonian that can produce wave packets with any desired momentum. 
The basic idea is to add linear diagonal terms to the elements of $H_{\rm PST}$, i.e.,
\begin{align}\label{eq:H_a_entries}
    J_n &= \sqrt{n(N+1-n)} & B_n &= an 
\end{align}
and we denote the new Hamiltonian by $H_a$. 
We want to compute the transition amplitude of the time evolution of $H_a$,
\begin{equation}\label{eq:H_a_amplitude}
     A(r,a,t) \equiv \mel{r}{ e^{-i H_a t}}{0} \ .
\end{equation}

\begin{figure*}[t]
    \centering
    \includegraphics[width =0.9\linewidth]{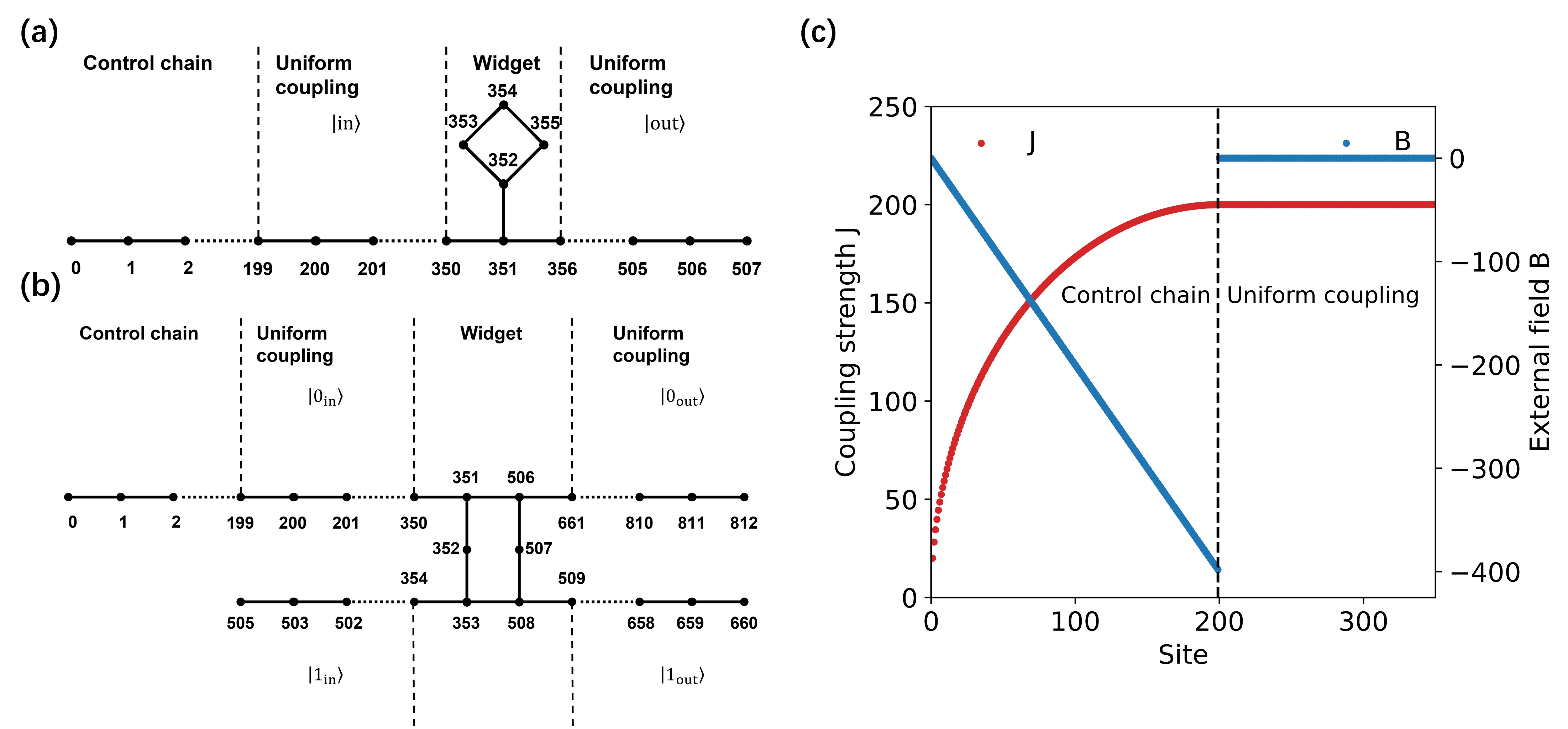}
    \caption{Settings of the scattering experiments for the two certain single-qubit gates. (a) Settings for $U_b$. (b) Settings for $U_c$. (c) The concrete data of the coupling strengths and the external fields on the control chain and the first uniform coupling chain.}
    \label{fig:settings}
\end{figure*}

According to Vinet and Zhedanov \cite{vinet2012construct}, for a general Jacobi matrix $H$ of the form~\eqref{eq:general_H}, it could be diagonalized into,
\begin{equation}
    H = PW \Lambda W^T P^T, 
\end{equation}
where
\begin{align}
\Lambda &= \diag(\lambda_0,\lambda_1, \dots,\lambda_N) \\
P &= 
\begin{pmatrix}
1  & 1 & \dots & 1 \\
\chi_1(\lambda_0) & \chi_1(\lambda_1) & \dots &\chi_1(\lambda_N) \\
\chi_2(\lambda_0) & \chi_2(\lambda_2) & \dots &\chi_2(\lambda_N)  \\
\vdots            & \vdots            & \vdots&\vdots             \\
\chi_N(\lambda_0) & \chi_N(\lambda_2) & \dots &\chi_N(\lambda_N)  \\ 
\end{pmatrix} \\
W &= \diag(\sqrt{\omega_0}, \sqrt{\omega_1}, \dots, \sqrt{\omega_N}) \ .
\end{align}
Here, $\lambda_n$ is the $n$-th eigenvalues of $H$; $\omega_n$ and the orthogonal polynomial $\chi_n(x)$ satisfy the following orthogonality relation,
\begin{equation}\label{eq:chi_orthogonality}
    \sum_{s=0}^N \omega_s \chi_n(\lambda_s) \chi_m(\lambda_s) = \delta_{mn} \ .
\end{equation}
Moreover, $\chi_n(x)$ satisfies the three-term recurrence relation,
\begin{equation}\label{eq:chi_recurrence}
    x \chi_n(x) =  J_n \chi_{n-1}(x) + B_n \chi_n(x) + J_{n+1} \chi_{n+1}(x), 
\end{equation}
with initial conditions $\chi_{-1}(x) = 0$ and $\chi_0(x) =1$.
Finally, note that $e^{-iHt} = P W e^{-i\Lambda t} W^T P^T$, and that
\begin{align}
    W^T P^T \ket{0} &= (\sqrt{\omega_0}, \cdots, \sqrt{\omega_N})^T \\
    W^T P^T \ket{r} &= (\sqrt{\omega_0} \chi_r(\lambda_0), \cdots, \sqrt{\omega_N} \chi_r(\lambda_N))^T \ .
\end{align}
This gives the expression for the transition amplitude of the time evolution operator $e^{-iHt}$,
\begin{equation}\label{eq: general_amp}
    \bra{r} e^{-i H t} \ket{0} = \sum_{n=0}^N \omega_n \chi_r(\lambda_n) e^{-i\lambda_n t} \ ,
\end{equation}
which holds for a general Jacobi matrix $H$.

To find an analytic expression for $A(r, a, t)$, one can first find the set of eigenvalues $\lambda_n$ of $H_a$, and then derive $\omega_n$ and $\chi_r(\lambda_n)$ using the orthogonality relation~\eqref{eq:chi_orthogonality} and the recurrence relation~\eqref{eq:chi_recurrence}.
Then, substituting them into Eq.~\eqref{eq: general_amp} gives 
\begin{eqnarray}\label{eq:A_a}
    A(r,a,t) = && \sqrt{\binom{N}{r}} q^{\frac{r}{2}} (1-q)^{\frac{N-r}{2}} e^{i r \theta(t,a)},
\end{eqnarray}
where $q \equiv 4b^2\sin^2\left(\frac{t}{2b}\right)$ and $b \equiv \sqrt{\frac{1}{a^2+4}}$; for the detailed derivation, we refer to Appendix~\ref{Appendix-A}.
$|A(r,a,t)|^2$ is the probability of finding the walker at site $r$ after an evolution time $t$ and satisfies the binomial distribution here.
The momentum of the wave packet is given by $\theta(t, a)$, which has the following expression,
\begin{equation}\label{eq:momentum}
    \theta(t,a)= 
    \begin{cases}
    -\tan^{-1}(ab \tan{(\frac{t}{2b})}) - \frac{\pi}{2}, \text{$t \in (0,b\pi]$} \\
    -\tan^{-1}(ab \tan{(\frac{t}{2b})}) + \frac{\pi}{2}, \text{$t \in (b\pi,2b\pi)$}\\
    \end{cases} .
\end{equation}

\begin{figure*}[t]
    \centering
    \includegraphics[width =\linewidth]{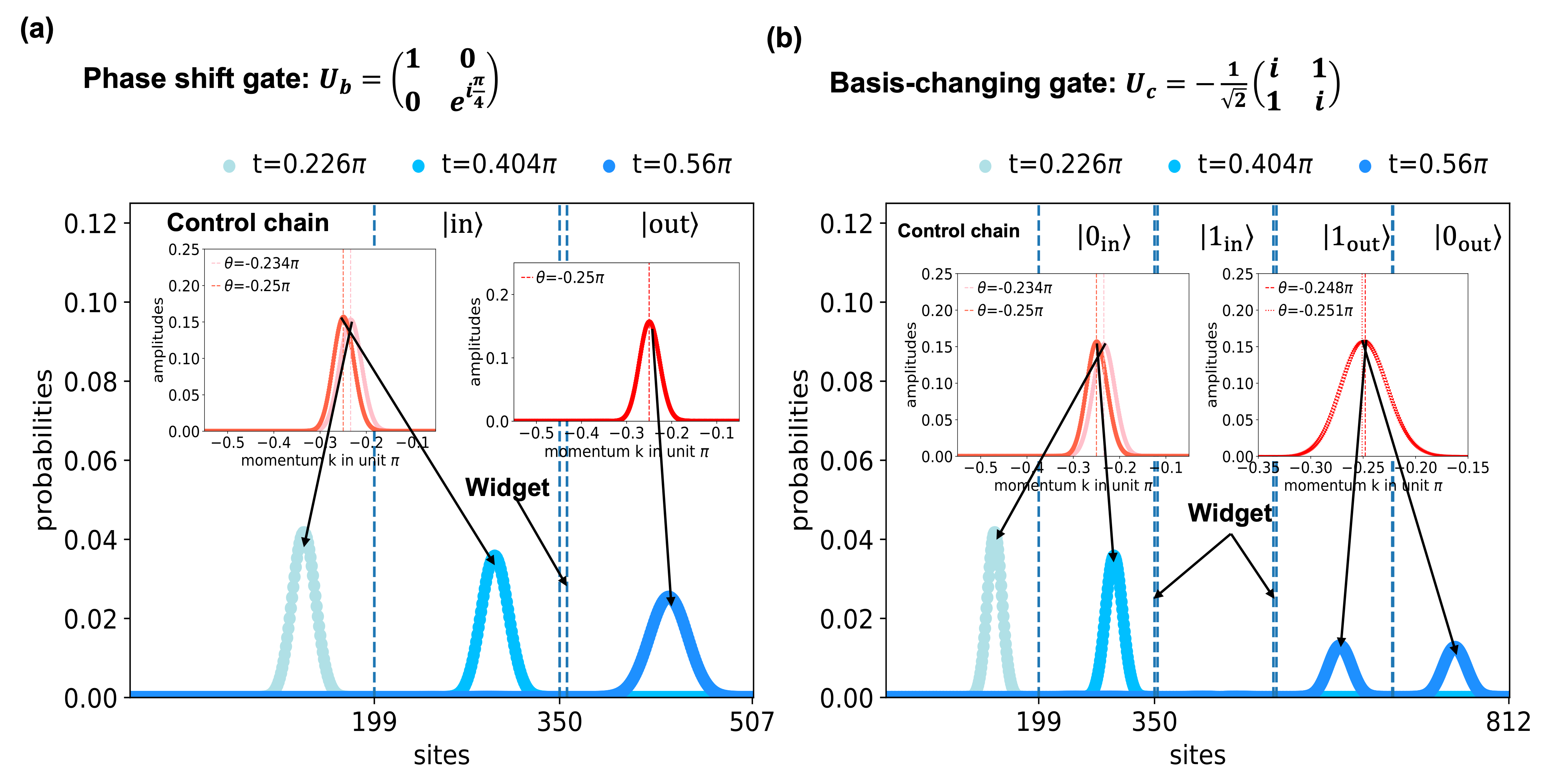}
    \caption{Time evolution processes including the wave packet preparation and scattering process in position space and corresponding momentum space for simulating $U_b$ and $U_c$.}
    \label{fig: scattering_results}
\end{figure*}

Specifically, choosing $ a = 0 $, $H_{\rm PST}$ can be recovered from Eq.~\eqref{eq:H_a_entries}
and the transition amplitude reduces to:
\begin{equation}
    \sqrt{\binom{N}{r}} (\sin{t})^r (\cos{t})^{N-r} e^{- i\frac{\pi}{2}r},
\end{equation}
in time $t \in (0, \pi/2]$, which is consistent with Eq.~\eqref{eq:H_PST_amplitude}.

Above, we obtain the explicit expression of the momentum as a function of the evolution time $t$ and the parameter $a$ of magnetic field.
Fig.~\ref{fig:momentum} gives visualization of this expression, which shows that one can obtain arbitrary momentum in $(-\pi,\pi]$ by varying $a$ and $t$.
When $ a > 0$ and $0 < t < \frac{2\pi}{\sqrt{a^2+4}}$, $\theta$ could cover the interval $(-\pi,-\frac{\pi}{2}) \cup (\frac{\pi}{2}, \pi]$ and when $ a<0 $, $\theta$ could take value from the interval $(-\frac{\pi}{2},\frac{\pi}{2})$.
When $a=0$, our method reduces to the original case of $H_{\rm PST}$, and hence $\theta$ can just be $\pm \frac{\pi}{2}$.
Also note that in all the cases, the lines are symmetric about $(b \pi, 0)$ with $b = \sqrt{\frac{1}{a^2+4}}$, which indicates that the state of the walker comes back to $\ket{0}$ when $t=2b\pi$ and completes a periodic motion.

\section{universal quantum gates}

As a demonstrative example, we apply our method to Childs' method for implementing universal quantum computation.
According to Childs~\cite{childs2009universal}, by scattering wave packets with the momentum of $-\frac{\pi}{4}$ with the widgets displayed in Fig.~\ref{fig:settings}~(a) and (b), two specific single-qubit gates could be realized,
\begin{align}
U_b &\equiv
\begin{pmatrix}
1 & 0  \\
0 & e^{i\frac{\pi}{4}} 
\end{pmatrix} &
U_c &\equiv -\frac{1}{\sqrt{2}}
\begin{pmatrix}
i & 1  \\
1 & i
\end{pmatrix} \ .
\end{align}
These two single-qubit gates, together with the CNOT gates, achieve universal quantum computation.

Below, we explain Childs' method in more detail.
The wire in Fig.~\ref{fig:settings}~(a) simulates the action of $U_b$ applied to $\ket{1}$, while a similar wire without the widget simulates the action of $U_b$ on $\ket{0}$.
After the control chain (i.e., quantum slide, used for generating wave packets from a single node indexed as 0), the wave packet will evolve into the state $\ket{\tin}$ with momentum $-\pi/4$, which is to guarantee that it can be transmitted through the widget perfectly (i.e., without reflection).
Note that $\ket{\tin}$ is not a static state; instead, it is a wave packet that will propagate forward.
Then, after the widget, the wave packet will gain a global phase $e^{i\pi/4}$, reminiscent of $U_b \ket{1}$; symbolically, we denote the wave packet after the widget as $\ket{\tout}$, which is also not static. 
Without the widget, the wave packet represented by $\ket{\tin}$ just propagates to become $\ket{\tout}$, without an extra global phase of $e^{i\pi/4}$; this simulates the action of $U_b$ applied to $\ket{0}$.
These two wires combine to simulate the action of $U_b$.
Similarly, in Fig.~\ref{fig:settings}~(b), the integrated part of the uniform coupling and the widget simulates the action of $U_c$.
As in the case of $U_b$, the wave packets represented by $\ket{0_{\tin}}$ and $\ket{1_{\tin}}$ need to be of momentum $-\pi/4$, so that they can be transmitted through the widget perfectly.
For our later numerical experiment, we connect the control chain to the wire corresponding to $\ket{0_{\tin}}$, in order to simulate the preparation of the logical $\ket{0}$ state.

Therefore, in Childs' method, an important component is to generate wave packets with momentum $-\pi/4$, which could be achieved by the Hamiltonian $H_a$.
As shown in Fig.~\ref{fig:momentum}, wave packets with the momentum of $-\frac{\pi}{4}$ can indeed be generated by choosing an appropriate $a$ from $(-\infty,0)$ (we choose $a = -2$ in the following) and controlling the corresponding evolution time $t$ accurately.

However, there are some technical issues to be solved in practice.
First, we need to stop the further evolution of the wave packet when its momentum reaches the value we want. 
Here, we adopt the idea of the quantum slide method, which is to ``cut'' the previously mentioned chain corresponding to $H_a$ in half as our control chain and connect an uniformly coupled chain behind it. 
Such ``cutting'' can be achieved by preparing a chain of length $N$ and setting the coupling strength of $H_a$ to be $J_n = \sqrt{n(2N-n)}$.
In this way, we virtually prepare a chain of length $2N$, and only take the first half as the control chain.
The uniform chain is to stably transmit the wave packet with the specific momentum for subsequent scattering with the widgets. 
According to Childs' theory, there should be uniformly coupled chains on both sides of the widgets, which is compatible with the quantum slide method.

Second, what we want is a wave packet with momentum of $-\pi/4$ on the uniformly coupled chain on the left side of the widgets rather than on the control chain.
However, the wave packet will not travel through the connection point smoothly.
On the one hand, we need to turn off the external fields on the control chain at an appropriate time, so that the wave packet will not reflect at the connection site.
On the other hand, the momentum of the wave packet will have a small shift, the magnitude of which depends on the time to turn off the field.
Therefore, the time to turn off the field is not exactly when the momentum of the wave packet reaches $-\frac{\pi}{4}$ on the control chain.
We need to choose a time that makes the wave packet possess a $-\frac{\pi}{4}$ momentum on the left uniform coupling chain after the transition.

After considering these issues, we perform numerical experiments with settings as detailed in Fig.~\ref{fig:settings}. 
For $U_b$ as shown in Fig.~\ref{fig:settings}~(a), it contains a control chain (nodes 0-199) to ``decelerate'' the wave packet (from an initial momentum that is always $-\frac{\pi}{2}$), two uniform coupling chains which are related to the input (nodes 200-350) and the output (nodes 356-507) of $U_b$ gate and a widget contributing a $\frac{\pi}{4}$ global phase shift to the incident wave packet with the momentum of $-\frac{\pi}{4}$.
Our simulation showed that the length of these uniform coupling chains does not have a big impact on the precision of the gates, as long as it is enough to accommodate the wave packets. 
But the length of the control chains will strongly affect the precision of the gates, which we will discuss later.
For $U_c$ as shown in Fig.~\ref{fig:settings}~(b), it differs from $U_b$ in the widget.
As for the length of the chains, including the control chain, inputs and outputs, they are all chosen to have the same length as the experiment of $U_b$ for convenience.
The coupling strengths and the external field of the control chain and the left uniform coupling chain are shown in Fig.~\ref{fig:settings}~(c).
The first 200 nodes as the control chain are effectively the first half of a 400-node chain which has a corresponding 400-by-400 $H_a$ with $a = -2$. 
The following 151 nodes as the input for the quantum gates are all uniformly coupled (the coupling strength equals that between the last two sites of the control chain) and with no external field. 
In addition, the remaining, including output chains and widgets, are all uniformly coupled and with zero external field.

The optimal time to turn off the external field under this specific setting is determined by repeated tests, and the value is $t=0.226\pi$ as shown in Fig.~\ref{fig: scattering_results}.
In this way, the momentum of the wave packets on the uniform coupling chains becomes $-0.25\pi$ after the momentum shifts in the transition point between two chains, as indicated by the snapshots at $t= 0.404\pi$.
After the scattering process in the widgets, the wave packet transmitted through the widget of $U_b$ has transmission probability $99.71\%$, which also represents the precision of the gate $U_b$.
As for $U_c$, the precision reaches $99.03\%$ and the wave packet split into two nearly equal parts as shown in Fig.~\ref{fig: scattering_results} (b).

\begin{figure}[t]
    \centering
    \includegraphics[width =\linewidth]{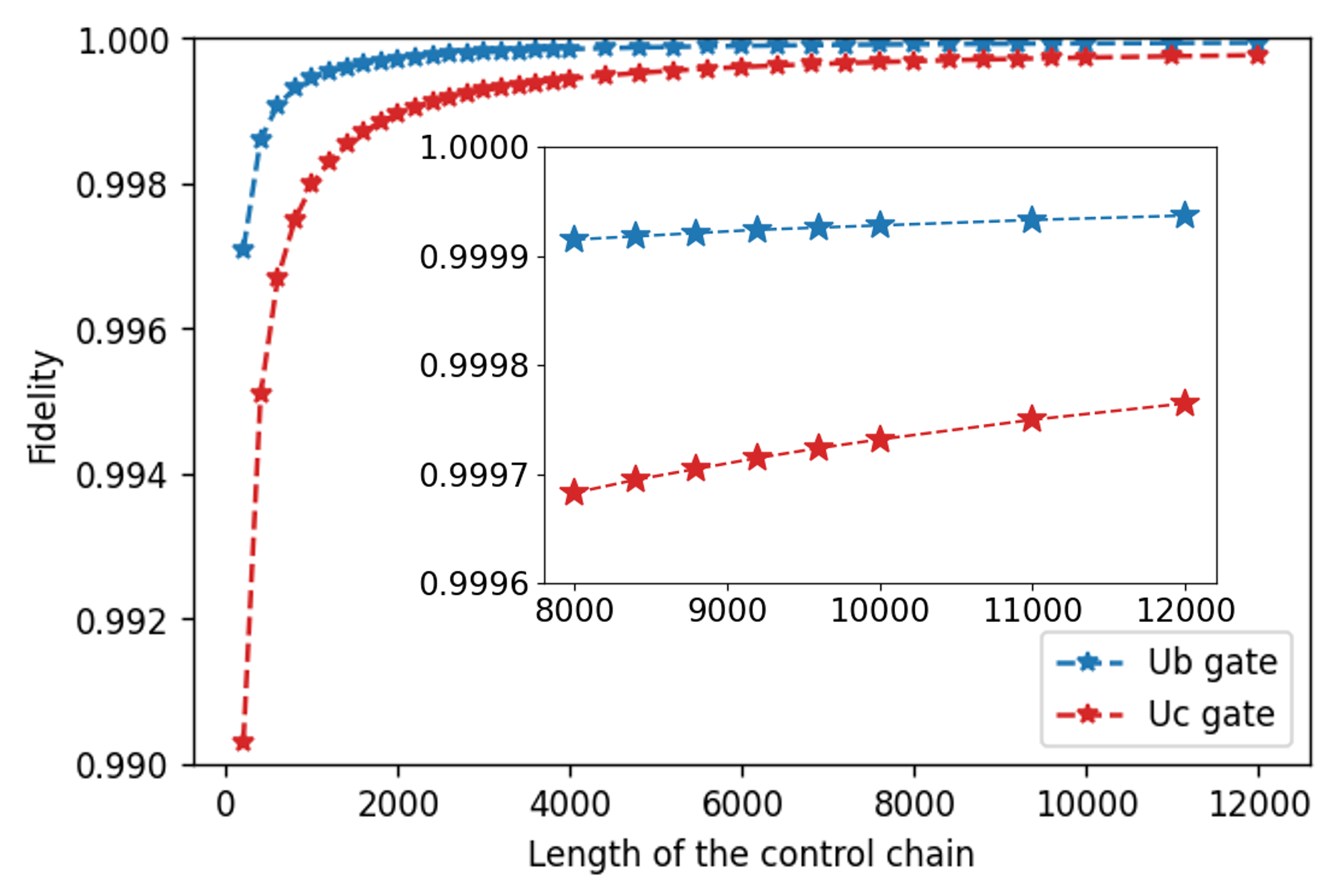}
    \caption{Fidelity of $U_b$ and $U_c$ by increasing the length of the control chain.}
    \label{fig:fidelity}
\end{figure}

Moreover, further numerical experiments show that higher precision can be achieved with longer control chain.
With a control chain consisting of thousands of nodes, the precision could reach more than 99.9\% or even 99.99\% as illustrated in Fig.~\ref{fig:fidelity}.
Here, for every increase of 200 nodes in the control chains, we increase the length of the uniform coupling chains by 25 to adapt to the wider wave packets. 
Additionally, in Appendix~\ref{Appendix-B}, we give a mathematical analysis of the influence of the length of the control chain on the fidelity and show that when the length is large enough, the fidelity could reach 1 asymptotically.

The successful experimental implementation of our proposed scheme relies on incorporating a linear external field to the quantum slide. An approach that holds promise for completing this task is the utilization of the femtosecond laser direct writing technique~\cite{crespi2013integrated,tang2018experimental, chaboyer2015tunable} in constructing a system of coupled waveguides. With the precise control of writing speed, this technique enables the introduction of a targeted constant detuning to the propagation constant of each waveguide, including various segments within a single waveguide~\cite{tang2022generating}. Since the propagation constants correspond to the diagonal terms of the system Hamiltonian, this manipulation provides an effective means of introducing a desired external field to each node of the quantum slide.

\section{Conclusion}
In this work, we propose the improved quantum slide method to generate wave packets with arbitrary momentum.
As a demonstrative example, we apply it to Childs' scheme to achieve universal quantum computing.
Our method extends the original quantum slide method via adding a linear external field to the original slide.
We derived an analytic expression of the momentum as a function of the evolution time $t$ and the parameter $a$ of the external field.
As a result, one can prepare wave packets of the desired momentum by controlling the two parameters.
This allows us to apply the improved quantum slide method to Childs' scheme for universal quantum computation, which requires incident wave packets with momentum $-\frac{\pi}{4}$ for later scattering processes.
Furthermore, we show how to achieve high-fidelity gates with limited length of the quantum slide (i.e., the control chain).  
The fidelity of the gates could be improved by increasing the length of the slide and could reach 1 asymptotically when the slide is long enough.
Therefore this method might become a powerful tool in the development of universal quantum computation. 
Lastly, our analysis leads us to suggest that the system of coupled waveguides, which has been fabricated utilizing the femtosecond laser direct writing technique, represents a potentially promising candidate for the successful realization of our proposed scheme.
Our work shall stimulate more development in the experimental technology in the near future.


\clearpage
\bibliography{ref}

\clearpage
\appendix
\section{Detailed derivation of the methodology}\label{Appendix-A}


In this Appendix, we give the details of deriving $A(r, a, t)$.
First, it is usually convenient to use the monic polynomials related to $\chi_n(x)$,
\begin{equation} \label{eq:relation_P_monicP}
    P_n(x) := J_1 J_2\dots J_n \chi_n(x) \ , 
\end{equation}
which has the recurrence relation,
\begin{equation}\label{eq:P_recurrence}
    x P_n(x) = J_n^2 P_{n-1}(x) + B_n P_n(x) + P_{n+1}(x) \ . 
\end{equation}

Next, we should have applied \eqref{eq: general_amp} to $H_a$,
but we first consider a related Hamiltonian which we denote as $H_p$ with:
\begin{align}\label{eq:Hp}
    J_n &= \sqrt{p(1-p)n(N+1-n)} & B_n &= (1-2p)n+pN \ .
\end{align}
We will first derive the transition amplitude $A(r,p,t)$ of the time evolution
$H_p$ because it is mathematically more convenient than $H_a$ in derivation. And then utilizing the relationship between $H_a$ and $H_p$, we can easily obtain $A(r,a,t)$ from $A(r,p,t)$. So we will apply \eqref{eq: general_amp} to $H_p$ to get $A(r,p,t)$ in the following.

First of all, $H_p$ has an interesting property that for whatever value of $p$ in interval $(0,1)$, its eigenvalues always observe this rule:
\begin{equation}\label{eq:eigenvalues_Hp}
    \lambda_n = n, n = 0,1,\dots, N. 
\end{equation}
Secondly, we need to find the corresponding orthogonal polynomial $\chi_n(x)$ of $H_p$. Here we consider the three-term recurrence relation of Krawchouk polynomials \cite{ismail2005classical,olver2010nist}
$K_n(x,p)$ ($p$ can be viewed as a constant temporarily):
\begin{eqnarray}\label{eq:recurrence_relation_kraw}
    -x K_n(x,p) = && p(N-n)K_{n+1}(x,p) -[p(N-n)+n(1-p)] \nonumber \\ && \times K_n(x,p) +n(1-p)K_{n-1}(x,p) ,
\end{eqnarray}
or alternatively the recurrence relation of the corresponding monic polynomials:
\begin{eqnarray}\label{eq:recurrence_relation_monic_kraw}
    x P_n(x,p) &&= np(1-p)(N+1-n)P_{n-1}(x,p)+ \nonumber \\ && [(1-2p)n + pN]P_n(x,p) + P_{n+1}(x,p) .
\end{eqnarray}
And the relations between the two polynomials are:
\begin{equation}\label{eq:monic_kraw}
    P_n(x,p) = (-N)_n p^n K_n(x,p) \ ,
\end{equation}
where $(-N)_n \equiv (-1)^n \binom{N}{n}$. By comparing \eqref{eq:recurrence_relation_monic_kraw}, \eqref{eq:P_recurrence} and \eqref{eq:Hp}, we can conclude that for $H_p$, its $P_n(x)$ is just the monic Krawchouk polynomial \eqref{eq:monic_kraw}.
Hence for $H_p$, its $\chi_n(x)$ can be found by \eqref{eq:relation_P_monicP}:
\begin{eqnarray}\label{eq:chi_nx}
    \chi_n(x) &=& \frac{1}{J_1 J_2\dots J_n} P_n(x,p) \nonumber \\
    &=& (-1)^n (p)^{\frac{n}{2}} (1-p)^{-\frac{n}{2}} \sqrt{\binom{N}{n}} K_n(x,p), 
\end{eqnarray}
with $J_1 J_2\dots J_n = \sqrt{\binom{N}{n}} (p(1-p))^{n/2}$.
And for Krawchouk polynomials, weight functions are in the following form:
\begin{equation}\label{eq:weight_functions}
    \omega_n = \binom{N}{n} p^n (1-p)^{N-n}. 
\end{equation}
And we can check that \eqref{eq:weight_functions} is also the $\omega_n$ of $H_p$ by substituting it and \eqref{eq:chi_nx} into \eqref{eq:chi_orthogonality}.

Now putting (\ref{eq:eigenvalues_Hp}), (\ref{eq:chi_nx}), (\ref{eq:weight_functions}) into (\ref{eq: general_amp}), the amplitudes of final states after evolution could be expressed as:
\begin{eqnarray}\label{eq:Hp_amp_1}
    A(r,p,t) &&= (-1)^r (p)^{\frac{r}{2}} \nonumber  (1-p)^{-\frac{r}{2}} \sqrt{\binom{N}{r}}\sum_{n=0}^N \binom{N}{n} p^n \nonumber \\
    && \times (1-p)^{N-n} K_r(n,p) e^{-int}.
\end{eqnarray}
From the self-duality of Krawchouk polynomials, i.e. $K_r(n,p) = K_n(r,p) (n,r = 0,1,\dots,N)$, then
\begin{eqnarray}\label{eq:Hp_amp_2}
    (\ref{eq:Hp_amp_1}) &&= (-1)^r (p)^{\frac{r}{2}} (1-p)^{N-\frac{r}{2}} \sqrt{\binom{N}{r}} (1- e^{-it} )^r\nonumber\\
    &&(1+\frac{p}{1-p} e^{-it})^{N-r}, 
\end{eqnarray}
where the generating functions of Krawchouk polynomials have been used:
\begin{equation}
    (1- \frac{1-p}{p} z)^r (1+z)^{N-r} = \sum_{n=0}^N  \binom{N}{n} K_n(r,p) z^n, 
\end{equation}
by letting $z =\frac{p}{1-p} e^{-it} $.\\
Then consider the following equations:
\begin{eqnarray}
    1 + c e^{-it}&& =  1 + c \cos{t} - i c \sin{t} = \sqrt{c^2 + 2c\cos{t} +1} e^{i \alpha},\nonumber \\
    && \alpha = -\arctan{\frac{c \sin{t}}{1 + c\cos{t}}}, 
\end{eqnarray}
and substitute it into (\ref{eq:Hp_amp_2}) by letting $c = -1, c = \frac{p}{1-p}$ respectively, then we get the final result of the amplitudes:
\begin{eqnarray}\label{eq:H_p_amplitude}
    A(r,p,t) && =\sqrt{\binom{N}{r}} [4p(1-p)\cos^2(\frac{t}{2})+(1-2p)^2]^{\frac{N-r}{2}}   \times \nonumber \\ 
    &&[4p(1-p)\sin^2(\frac{t}{2})]^{\frac{r}{2}} e^{i r\theta(t,p)} e^{i N \theta_g(t,p)},
\end{eqnarray}
with
\begin{equation}\label{eq:momentum_p}
    \theta(t,p) = 
    \begin{cases}
    -\arctan((1-2p) \tan{(\frac{t}{2})}) - \frac{\pi}{2}, \text{$t \in (0,\pi]$ } \\
    -\arctan((1-2p) \tan{(\frac{t}{2})}) + \frac{\pi}{2}, \text{$t \in (\pi,2\pi)$}\\
    \end{cases} ,
\end{equation}
\begin{equation}
  \theta_g(t,p) = \arctan{(\frac{p\sin{t}}{1-p+p\cos{t}})}.
\end{equation}

Now back to case of  $H_a$, we consider:
\begin{align}
    a &= \frac{1-2p}{\sqrt{p(1-p)}},
\end{align}
and for $p \in (0,1)$, we have $a \in (-\infty,\infty)$.
Conversely,
\begin{equation}\label{eq:p(a)}
    p = 
    \begin{cases}
    \frac{1}{2}(1+\sqrt{\frac{a^2}{a^2+4}}), \text{$a <= 0 $} \\
    \frac{1}{2}(1-\sqrt{\frac{a^2}{a^2+4}}), \text{$a > 0$} \\
    \end{cases}.
\end{equation}
Then, ignoring some unimportant constant diagonal terms, which only bring some global phases to the final state, $H_a$ and $H_p$ can be related by the following formula:
\begin{equation}\label{eq:relation_Ha_Hp}
    H_a = \frac{H_p}{\sqrt{p(1-p)}}.
\end{equation}
Hence, up to some global phases, 
\begin{align} \label{eq:Ha_amplitudes}
     \mel{r}{ e^{-i H_a t}}{0} &= \mel{r}{ e^{-i H_p \frac{t}{\sqrt{p(1-p)}}}}{0} \notag \\ 
     &= A(r,p,\frac{t}{\sqrt{p(1-p)}})
\end{align}
Lastly put the Eq.~\eqref{eq:H_p_amplitude} and \eqref{eq:p(a)} into Eq.~(\ref{eq:Ha_amplitudes}), we have the final results shown in the equations (\ref{eq:A_a})-(\ref{eq:momentum}).

\section{Analytical transmission probability of the wave packets}\label{Appendix-B}
Considering the wave packets with the amplitudes shown in \eqref{eq:A_a},
when N is large enough, we can use the binomial-Gaussian approximation:
\begin{equation}
    A(r,a,t) \approx \frac{1}{(2\pi \sigma^2)^{\frac{1}{4}}} e^{-\frac{(r-Nq)^2}{4\sigma^2}} e^{i\theta r}
\end{equation}
with $\sigma^2 = Nq(1-q)$.
Then take a Fourier Transformation to the amplitudes, we obtain the amplitudes in momentum space:
\begin{equation}
    \widetilde{A}(k,a,t) \approx \frac{\sqrt{2 \sigma}}{(2\pi)^{\frac{1}{4}} } e^{-\sigma^2 (k + \theta)^2} e^{i Nq(k + \theta)},
\end{equation}
hence the probability distribution in the momentum space which is also a Gaussian distribution with respect to $k$:
\begin{equation}
    |\widetilde{A}|^2(k,a,t) \approx \frac{1}{\sqrt{2\pi} \sigma_k} e^{-(k - \theta)^2/2 \sigma_k^2} 
\end{equation}
with the standard deviation $\sigma_k = \frac{1}{2\sigma}$.
And we know the transmission probability of the plane waves to the widgets from \cite{childs2009universal}, which we denote by $T^{(b)}(k)$ and $T^{(c)}(k)$; for instance,
\begin{equation}
    T^{(b)}(k) = \frac{64}{64+\cos^2{2k}\csc^6{k}\sec^2{k}}.
\end{equation}
for k $\in (-\pi,0)$.

Then the transmission probability for the Gaussian wave packets could be obtained as the expectation values of $T^{(b)}(k)$ and $T^{(c)}(k)$ in the probability distribution in the corresponding momentum space. So we can calculate the transmission probability of the Gaussian wave packets as a function of $\theta$ to the widget b via the following integral:
\begin{align}\label{eq:TGb}
    T_G^{(b)}(\theta) &=  \int_{-\pi}^0 |\widetilde{A}|^2(k,a,t) T^{(b)}(k) dk \notag \\
    =\int_{-\pi}^0 & \frac{1}{\sqrt{2\pi} \sigma_k} e^{-(k - \theta)^2/2 \sigma_k^2} \frac{64}{64+\cos^2{2k}\csc^6{k}\sec^2{k}} dk.
\end{align}
When $N \rightarrow \infty$, so $\sigma_k \rightarrow 0$, then the Gaussian distribution would approximate a Dirac delta function:
\begin{equation}
    \lim_{\sigma_k \rightarrow 0} \frac{1}{\sqrt{2\pi} \sigma_k} e^{-(k - \theta)^2/2 \sigma_k^2}  = \delta(k-\theta) .
\end{equation}
So \eqref{eq:TGb} can just reduce to the expression of a plane wave with the momentum $\theta$:
\begin{equation}
    T_G^{(b)}(\theta) =  \frac{64}{64+\cos^2{2\theta}\csc^6{\theta}\sec^2{\theta}} = T^{(b)}(\theta),
\end{equation}
which could reach 1 when $\theta =- \frac{\pi}{4}$; and the same for $T_G^{(c)}(\theta)$.
\end{document}